\documentclass[twocolumn,prd,aps,amsmath,amssymb,epsfigure]{revtex4}

\usepackage{epsfig}
\usepackage{graphicx}
\usepackage{dcolumn}
\usepackage{bm}

\begin{document}
\title{Tetracritical behavior in strongly interacting theories}
 \author{Francesco {\sc Sannino}}
 \email{francesco.sannino@nbi.dk}
\affiliation{NORDITA \& The Niels Bohr Institute, Blegdamsvej
  17, DK-2100 Copenhagen \O, Denmark }
 \author{Kimmo {\sc Tuominen}}\email{kimmo.tuominen@phys.jyu.fi}
 \affiliation{Department of Physics,  P.O. Box 35,
FIN-40014 University of Jyv\"askyl\"a, Finland\\
Helsinki Institute of Physics, P.O. Box 64, FIN-00014 University
of Helsinki, Finland }
\date{March 2004}

\begin{abstract}
We suggest a tetracritical fixed point to naturally occur in
strongly interacting theories. As a fundamental example we analyze
the temperature--quark chemical potential phase diagram of QCD
with fermions in the adjoint representation of the gauge group
(i.e. adjoint QCD). Here we show that such a non trivial
multicritical point exists and is due to the interplay between the
spontaneous breaking of a global $U(1)$ symmetry and the center
group symmetry associated to confinement. Our results demonstrate
that taking confinement into account is essential for
understanding the critical behavior as well as the full structure
of the phase diagram of adjoint QCD. This is in contrast to
ordinary QCD where the center group symmetry associated to
confinement is explicitly broken when the quarks are part of the
theory.
\end{abstract}

\maketitle

\section{Introduction}
Phase diagrams for strongly interacting theories are a topic of
past and current interest \cite{Brown:dm}, and the relation
between deconfinement and chiral symmetry restoration poses a
continuous challenge. In ordinary QCD these problems have been
intensively addressed via computer simulations \cite{lattice}. By
investigating such a relation in different strongly interacting
theories one gains insight on the ordinary QCD dynamics as well.
We recall, that the order parameter for deconfinement is the
Polyakov loop \cite{Polyakov:vu}, while the one for chiral
symmetry restoration is the quark condensate. The representation
of the matter with respect to the gauge group is known to play a
relevant role in the deconfining dynamics. Much attention in the
literature has been given to ordinary QCD with two or three
flavors. The presence of quarks in the fundamental representation
breaks the center group symmetry explicitly, and for massless
quarks only the chiral phase transition remains well defined. The
latter is then expected to drive the critical behavior
\cite{Mocsy:2003qw}. At nonzero and large quark masses the issue
of which transition, i.e. deconfining or chiral symmetry
restoring, dominates, becomes a non perturbative problem which
only lattice computations can currently solve.

The situation becomes clearer, at least in principle, for fermions
in the adjoint representation of the gauge group. Here one has two
well defined and independent order parameters, since the center
group symmetry remains intact in the presence of the fermions.
Lattice data seems to confirm the independence of the forces
driving independently the chiral and deconfining phase transition
both for two and three colors \cite{Karsch:1998qj,Kogut:1985xa}.

However, when two or more orders compete the resulting phase
diagram is expected to have a very interesting and rich structure
due to the possibility of multicritical behavior. This arises at
the intersection of critical lines characterized by different
order parameters. Our interest is in the case of two order
parameters. If the transition at the multicritical point is
continuous, then either bicritical or tetracritical behavior can
occur. Bicritical behavior occurs if a first-order line
originating from the multicritical point separates two different
ordered phases, each separated from the disordered phase by a line
of continuous transitions beginning from the multicritical point.
Tetracritical behavior on the other hand occurs if there exists a
mixed phase in which both types of ordering coexist, and which is
bounded by two critical lines meeting at the multicritical point.
It is also possible that the phase transition at the multicritical
point is of first order. This case is similar to the bicritical
one, with the distinction that the two lines separating the
disordered phase from the ordered ones, start from the
multicritical point as first order lines and then turn to second
order lines at tricritical points. A typical condensed matter
example of multicritical behavior is the phase diagram of
anisotropic antiferromagnets in a uniform magnetic field parallel
to the anisotropy axis \cite{Fisher2}. Further examples include
$^4$He \cite{Fisher1} and high-$T_c$ superconductors \cite{SC}.
Also, it has been suggested that a multricritical behavior might
emerge in the phase diagram of hadronic matter at finite baryon
chemical potential \cite{tetraQCD}. {}For two colors a
tetracritical behavior induced by a possible competition between a
diquark and a quark-antiquark phase has been investigated in
\cite{Vanderheyden:2001gx}.

In this paper we show that strongly interacting gauge theories
with fermions in the adjoint representation may very naturally
display a tetracritical behavior. {}Interestingly, the two
competing orders we will consider are confinement and chiral
symmetry. The critical behavior arising from two competing orders
has a long history. Investigations in anisotropic magnetic systems
were carried out at the mean field level in \cite{Fisher1}, and
subsequently in \cite{Fisher2} to first order in $\epsilon=4-D$,
where $D$ is the dimension of spacetime. More recently the
analysis has been carried up to order $\mathcal{O}(\epsilon^5)$ in
the $\epsilon$-expansion \cite{Calabrese:2003ia}.

In this work we propose that a non trivial multicritical point
exists in the temperature--quark chemical potential phase diagram
of QCD with fermions in the adjoint representation of the gauge
group (i.e. adjoint QCD). The two competing orders are chiral
symmetry and confinement. Our results suggest that taking
confinement into account is essential for understanding the
critical behavior as well as the full structure of the phase
diagram of adjoint QCD. This is in contrast to ordinary QCD where
the center group symmetry associated to confinement is explicitly
broken when the quarks are part of the theory.

In section \ref{multicritical} we briefly review the basic
classification of the multicritical points
\cite{Fisher2,Fisher1,Calabrese:2003ia} relevant for our
discussion. In section \ref{temperature} we study Yang-Mills
theories with fermions in the adjoint representation of the gauge
group at  temperature. Here we discuss the critical
behavior in the hypothetical case in which chiral symmetry and
confinement compete for order. The early lattice work
 \cite{Kogut:1985xa} seems to exclude the presence
of multicritical points. Nevertheless, we find instructive to
discuss this regime.

We then introduce, in section \ref{finitemu}, a nonzero quark
chemical potential for one Dirac flavor in the adjoint
representation of two colors. Then we proceed to show that a
multicritical point is quite likely to occur in the
temperature--chemical potential phase diagram. The two orders
correspond to the $Z_2$ symmetry (i.e. $O(1)$) and the $U(1)\sim
O(2)$, respectively. $Z_2$ is the center group symmetry associated
with confinement, while $O(2)$ is the baryon number which
spontaneously breaks due to the formation of diquark condensates.
Some analogous theories have been investigated directly via
lattice simulations \cite{Hands:2001ee}, and within the chiral
perturbation theory approach
\cite{Kogut:2000ek,{Splittorff:2002xn}}. We show that the
interplay between the two order parameters substantially affects
the phase diagram.

The multicritical point is predicted to be in the  $O(3)$
Heisenberg universality class, according to the classification in
\cite{Fisher2}, if the fixed point analysis is performed at one
loop in the $\epsilon=4-D$ expansion \cite{Fisher2}. If higher
orders are considered the fixed point is predicted to be a biconical
tetracritical point \cite{Calabrese:2003ia}. We finally suggest
possible applications of our results to QCD with fermions in the
fundamental representation of the gauge group.

\section{Classification of Multicritical
Points}\label{multicritical}

In this paper we will argue that certain strongly interacting
theories naturally lead to phase transitions, in the
temperature--quark chemical potential plane, possessing
multicritical points. The novelty is in the fact that this
multicritical behavior is a result of the interplay of
deconfinement and global symmetry breaking.

One of the most remarkable features of continuous phase
transitions is their universal character. There is, indeed, a rich
variety of systems which exhibit the same identical critical
behavior. When possible, it is convenient to introduce order
parameters to describe the phase transition. In our case we will
have two order parameters: one associated to deconfinement and the
other to a global symmetry. Note, that even though we start from a
fermionic theory, near the critical point of interest the relevant
effective degrees of freedom are bosonic, and are naturally
identified with the physical fluctuations of the order parameters.
This is the standard approach related to the study of phase
transitions.

Before moving to the theories of interest to us, we introduce in
this section the relevant definitions and the classification of
the multricritical behaviors emerging when two order parameters
compete for order. We will keep the discussion general.

Following \cite{Fisher2} and \cite{Calabrese:2003ia} when we have
two order parameters, $\ell$ and $\sigma$, which compete with
symmetries $O(N_1)$ and $O(N_2)$, respectively, one can write the
effective theory symmetric under $O(N_1)\oplus O(N_2)$. Up to
quartic terms the effective theory containing both order
parameters in $D$ Euclidean dimensions is:
\begin{eqnarray}
{\mathcal{L}} &=&
\frac{1}{2}(\partial_\mu\ell)^2+\frac{1}{2}(\partial_\mu\sigma)^2
+\frac{1}{2}m_\ell^2\ell^2+\frac{1}{2}m_\sigma^2\sigma^2\nonumber
\\
& & +\frac{\lambda}{4!}(\ell^2)^2 + \frac{g_4}{4!}(\sigma^2)^2 +
\frac{g_2}{4}\ell^2\sigma^2. \label{2otheory}
\end{eqnarray}
Here $\ell^2=\sum_{n=1}^{N_1} \ell_n^2$ and
$\sigma^2=\sum_{m=1}^{N_2} \sigma_m^2$. It is possible that for a
certain value of the physical parameters, and in D=3, the
correlation lengths of the two order parameters diverge
simultaneously yielding a multicritical point. At such a point the
critical behavior can be determined by tuning the parameters
$m_\ell^2$ and $m_\sigma^2$ to their critical values and studying
the stable fixed points of the renormalization group flow.

\subsection{Fixed points and critical behavior at one loop}
A first order analysis in the $\epsilon$-expansion \cite{Fisher2}
for the theory (\ref{2otheory}) at multicritical point shows, that
six distinct fixed points exist. Four of them have $g_2=0$, and
three of these, namely the gaussian, $O(N_1)$ and $O(N_2)$
symmetric ones, are always unstable against the perturbations away
from the $g_2=0$-plane, while the fourth one is stable for
sufficiently large values of $N_1$ and $N_2$. Since $g_2=0$, the
two fields behave independently and this stable fixed point is
termed {\it decoupled} fixed point. The other two stable fixed
points lie at nonzero $g_2$. First of them is called the
Heisenberg $O(N_1+N_2)$ fixed point, due to enhanced symmetry, and
the second one is called the {\it biconical} fixed point. The
fixed points can be determined by computing the zeros of the beta
functions of the theory which at one loop are:
\begin{eqnarray}
\beta(\lambda)&=&\frac{\lambda}{6}\frac{(N_1+8)}{8\pi^2}+3\frac{g_2^2}{2}\frac{N_2}{8
\pi^2} - \lambda \epsilon\ , \nonumber \\
\beta(g_4)&=&\frac{g_4}{6}\frac{(N_2+8)}{8\pi^2}+3\frac{g_2^2}{2}\frac{N_1}{8\pi^2}
-
\lambda \epsilon \ , \nonumber \\
\beta(g_2)
&=&\frac{\lambda\,g_2}{6}\frac{(N_1+2)}{8\pi^2}+\frac{g_4\,g_2}{6}\frac{(N_2+2)}{8\pi^
2} + 2\frac{g^2_2}{8\pi^2}-g_2\epsilon \nonumber \ .
\end{eqnarray}
The stability of a generic fixed point is ensured if the matrix
\begin{eqnarray}
\omega_{ij}=\left.\frac{\partial \beta_i}{\partial
g_i}\right|_{g^{\ast}} \ ,
\end{eqnarray}
evaluated at the fixed point $g^{\ast}$ has real and positive
eigenvalues. Our results agree with the ones in \cite{Fisher2}.

The nature of the multicritical point is determined by the sign of
the quantity $\lambda g_4-g_2^2/9$ \cite{Fisher1}. This constraint
simply tells us, at the level of the effective Lagrangian, if the
phase displaying two orders (i.e. non vanishing condensates for
both order parameters) has a higher or lower free energy with
respect to the phases in which one of the condensates vanishes
\cite{Fisher1}. If the sign is positive we expect a tetracritical
behavior. {}For the negative sign the phase with two orders has
higher free energy than the phases with only partial order. In
this case a simultaneous existence of two orders is unstable and a
jump between the phases with partial orders occurs. In the latter
case we expect a bicritical behavior.

Decoupled and biconical stable fixed points mentioned above
satisfy the criterion of tetracriticality, $\lambda g_4>g_2^2/9$,
at the critical point, while for the fixed point corresponding to
the isotropic $N_1+N_2$-vector model can, interestingly, be either
bicritical or tetracritical \cite{Calabrese:2003ia}. This
possibility arises due to the presence of a dangerous irrelevant
variable \cite{Bruce}.

Defining $n=N_1+N_2$, the low order $\epsilon$-expansion
calculation shows that for $n<4$ the critical behavior is due to
the stable fixed point corresponding to an isotropic
$O(n)$-Heisenberg model. As $n$ increases, the biconical fixed
point becomes stable and yields a new tetracritical behavior.
Finally, for large $n$, namely for $N_1 N_2+2N_1+2N_2\ge 32$, the
stable fixed point is the decoupled one, which leads to the
tetracritical behavior in which the two fields do not affect each
other.

We first summarize in table \ref{exponents} the generic results
which were obtained in the $\mathcal{O}(\epsilon)$ calculation,
and then we discuss them in the context of strong interactions.
\begin{table}[th]
\begin{tabular}{l||c|c|c}
  FP & $n=N_1+N_2$ & $\nu_1$, $\nu_2$ \\
  \hline & & \\
  Decoupled & $n\ge 10$ & $\frac{1}{2}+\frac{N_2+2}{N_2+8}\frac{\epsilon}{4}$,
$\frac{1}{2}+\frac{\epsilon}{12}$  \\
            & $(N_1=1)$ & \\ & & \\
  Biconical &$4\le n < 10$ &  \\
          & $(N_1=1)$ & \\
          & $N_2=3$ & $0.5+0.1250\epsilon$, $0.5+0.0417\epsilon$ \\
          & $N_2=4$ & $0.5+0.1336\epsilon$, $0.5+0.0560\epsilon$ \\
          & $N_2=5$ & $0.5+0.1403\epsilon$,
          $0.5+0.0667\epsilon$ \\
          & $N_2=6$ & $0.5+0.1460\epsilon$,
          $0.5+0.0741\epsilon$ \\
          & $N_2=7$ & $0.5+0.1515\epsilon$,
          $0.5+0.0789\epsilon$ \\
          & $N_2=8$ & $0.5+0.1568\epsilon$,
          $0.5+0.0816\epsilon$ \\
& & \\
  Heisenberg & $n<4$ & $\frac{1}{2}+\frac{n+2}{n+8}\frac{\epsilon}{4}$,
$\frac{1}{2}+\frac{1}{n+8}\frac{\epsilon}{2}$ \\
 & & \\
\end{tabular}
\caption{The critical exponents of the correlation length at the
various multicritical points of $O(N_1=1)\oplus O(N_2)$ theory.}
\label{exponents}
\end{table}
Each of these fixed points have interesting specific properties:
{}For the Heisenberg fixed point the symmetry is enhanced from
$O(N_1)\oplus O(N_2)$ to $O(N_1+N_2)$, and the theory
(\ref{2otheory}) becomes that of isotropic Heisenberg
$N_1+N_2$-component model, as can be seen by inspecting the
lagrangian (\ref{2otheory}) at the fixed point given by
$\lambda=g_4=3g_2$. The critical exponents $\nu_1$ and $\nu_2$
quoted in the table \ref{exponents} are defined in terms of the
eigenvalues $\lambda_\ell$ and $\lambda_\sigma$ of the
corresponding relevant variables $m_\ell^2$ and $m_\sigma^2$ in
the linearized renormalization group recursion relations as:
\begin{eqnarray}
\nu_1 &=& \frac{1}{\lambda_\ell}\ , \qquad  \nu_2 =
\frac{1}{\lambda_\sigma}.
\end{eqnarray}
These describe the divergence of the correlation lengths as a
function of suitable scaling fields, for example the reduced
temperature $t=T/T_c-1$ and, say a new scaling field $g$. The
latter can be a magnetic field, a quark mass parameter etc.

The effect of the perturbation controlled by $g$ is generally
captured by the crossover exponent $\phi$ defined through the
usual scaling formula for e.g. correlation length
\begin{eqnarray} \xi(T,g)\sim t^{-\nu}F(g/t^\phi),
\end{eqnarray}
and similarly for other thermodynamical quantities. Here
$\nu=\nu_1$ corresponds to $g=0$ case and $\phi=\nu/\nu_2$. The
crossover scaling function $F(z)$ is finite at $z=0$, but has
divergences at specific points and these divergences then modify
the $g=0$ behavior $\sim t^{-\nu}$.

Considering magnetic systems as an example, for $n<4$ at
$\mathcal{O}(\epsilon)$, the crossover corresponds to the weakly
anisotropic $n$-vector model, where the anisotropy is given by the
term $\sim g\ell^2\sigma^2$ in the isotropic Hamiltonian. In other
words, $\nu_1$ describes the divergence of the correlation length
as $\xi\sim |t|^{-\nu_1}$, where $t$ is the reduced temperature,
while the exponent $\nu_2$ describes the divergence of the
correlation length in the anisotropy $g$ as $\xi\sim g^{-\nu_1}$
when $g\rightarrow 0$. The crossover exponent in this case is
given by $\phi=1+\frac{n\epsilon}{2(n+8)}$.

The decoupled fixed point describes a system consisting of
effectively independent $N_1$- and $N_2$-component Heisenberg
subsystems, and therefore the critical indices are the ones of the
two independent Heisenberg subsystems and are in that respect
trivial. Interestingly, though, the total scaling will break,
since a single scaling function cannot properly describe the
asymptotic free energy when $N_1\neq N_2$. Finally, the biconical
fixed point features completely new critical exponents. However,
since they are numerically very close to the corresponding
Heinseberg ones, they may be hard to distinguish experimentally.

\subsection{Results from higher order computations}

It is important to note that the numbers quoted in table
\ref{exponents} are a result of a first order calculation in
$\epsilon$. Furthermore, these results must be extrapolated to
$\epsilon=1$ to be applicable. However, past experience has shown
that even in this limit, the $\epsilon$ expansion describes the
fixed point physics surprisingly well. Already the
$\mathcal{O}(\epsilon)$ results show that as a function of $n$ the
critical behavior in the case of two competing orders leads to a
rich spectrum of possibilities. However in the present case higher
order contributions are relevant. For the case of $O(N_1)\oplus
O(N_2)$ theory, a remarkable $\mathcal{O}(\epsilon^5)$ calculation
exists \cite{Calabrese:2003ia}, and we will briefly discuss the
improvements for the critical exponents in what follows. First,
however, let us note that the higher orders also lead to important
changes in the domains of stability of the fixed points in the
$(N_1,N_2)$-plane. The $\mathcal{O}(\epsilon^3)$ results for the
Heisenberg fixed point \cite{Ketley} lead to the stability for
$N_1+N_2<4-2\epsilon+\frac{5}{12}(6\zeta(3)-1)+\mathcal{O}(\epsilon^3)$.

A calculation to $\mathcal{O}(\epsilon^5)$ further narrows the
domains of stability for the fixed points: The $O(n)$-Heisenberg
fixed point is stable only for $n=2$, i.e. only in the case of two
intersecting Ising lines. Then, for $n=3$ the stable fixed point
is the biconical one, and the decoupled fixed point is stable for
all $n\ge 4$ with any values of $N_1$ and $N_2$. For further
details we refer to the existing literature
\cite{Fisher2,Calabrese:2003ia}.

{}Since the domain of stability of the Heisenberg fixed point
shrinks down to $n=2$ it will not play a role in our strong
interaction examples. The biconical fixed point is stable for
$n=3$. We will see that this fixed point will be relevant for our
investigations. {}For all of the other combinations of $N_1$ and
$N_2$ such that $N_1+N_2\ge 4$, the stable fixed point is the
decoupled one with well known independent $O(N_1)$ and $O(N_2)$
exponents. Therefore, to conclude this section, let us state the
high order values for the critical exponents relative to the
biconical fixed point at $\epsilon=1$. Using the general
definitions $\nu=\nu_1$ and $\phi=\nu/\nu_2$, the numerical
Pad\'{e}--Borel resummed $\mathcal{O}(\epsilon^5)$ values for the
biconical exponents at $N_1=1$ and $N_2=2$ are: $\nu_B=0.70(3)$,
$\phi_B=1.25(1)$. As already mentioned, these are very close to
the corresponding Heisenberg $O(3)$ exponents: $\nu_H=0.7045(55)$,
$\phi_H=1.260(11)$ at $\mathcal{O}(\epsilon^5)$ \cite{ZJ}.

Away from the tetracritical points the second order lines have
independent critical behaviors and the two order parameters do not
compete.

We have now the basic terminology and tools to analyze and make
predictions for strongly interacting theories exhibiting
multicritical behavior.
\section{Finite Temperature Adjoint QCD}
\label{temperature}

Let us now turn to the possibility of tetracritical behavior in
the theories of strong interactions. To be specific, consider two
color QCD with $N_f\le 2$ massless Dirac flavors in the adjoint
representation of the gauge group. One of the main motivations for
studying the phase diagram of gauge theories with fermions in the
adjoint representation (adjoint QCD) is that, contrary to ordinary
QCD, in adjoint QCD there is a well defined symmetry associated to
confinement. The symmetry is identified with the center of the
gauge group which for a generic $SU(N)$ gauge theory is $Z_N$.
Here we consider explicitly the case $N=2$. The breaking of this
symmetry is monitored by the expectation value of the Polyakov
loop \cite{Polyakov:vu} which is the order parameter of the
theory.

Besides the center group symmetry, and in absence of quark masses,
adjoint QCD possesses a global quantum symmetry which for $N_f$
Dirac fermions is $SU(2N_f)$ \footnote{The classical symmetry is
$U(2N_f)=SU(2N_f) \times U_A(1)$. The axial global $U(1)_A$ is
explicitly broken by the Adler-Bell-Jackiw anomaly.}. The fact
that the symmetry group here is $SU(2N_f)$ rather than $SU(N_f)
\times SU(N_f) \times U(1)$ is due to the fact that the fermions
belong to a real representation of the gauge group. We note that
the ordinary baryon number is one of the diagonal generators of
$SU(2N_f)$. If a democratic Dirac mass term is added into the theory,
$SU(2N_f)$ breaks explicitly to $SU(N_f)\times U(1)$, with $U(1)$
the baryon number of the theory. In this section we consider the
massless limit, but note that the introduction of a small mass term for
the fermions in the theory can be introduced and studied in a
straightforward way. At low temperatures the global symmetry is
expected to break to the maximum diagonal subgroup $O(2N_f)$
leaving behind a number of goldstone bosons, some of them charged
under the ordinary baryon number. We will collectively refer to
the goldstone bosons as pions and will also use, at times, chiral
symmetry to indicate the global symmetry of the theory. In the
next section, and for the specific case of two colors and one
Dirac flavor, we will work out in detail the global
symmetry properties for massless and massive fermions. We will
also discuss the breaking patterns of the global symmetry, and
consider old and new arguments supporting these patterns.
At high temperatures it is natural to expect a global
symmetry restoration. Such a global symmetry restoration is also
termed, at times, chiral symmetry restoration.

We now naturally have two well defined order parameters: The
Polyakov loop and the fermion condensate. It is interesting to
consider the possibility that they may compete for order when
considering a temperature driven phase transition. The hope being,
as already mentioned in the introduction, that by studying
strongly interacting theories such as adjoint QCD, one might shed
light on ordinary QCD.

Having outlined the general behavior, symmetries and defined the
order parameters it is now natural to use the results and
methodology presented in the previous section to make predictions
for the critical exponents related to the phase transitions of
adjoint QCD.

{}For two colors the center group is $Z_2$, which is equivalent to
a $O(1)$ symmetry and the associated order parameter is denoted by
$\ell$. The flavor groups $SU(4)$ for $N_f=2$ and $SU(2)$ for
$N_f=1$ are locally isomorphic respectively to $O(6)$ and $O(3)$,
and the order parameter with such symmetry is denoted by $\sigma$.

Here the results of \cite{Calabrese:2003ia}, denoted by
$O(1)\oplus O(6)$ and $O(1)\oplus O(3)$, are directly applicable.
The first phase diagram we draw is the one in which the
temperature drives the phase transition at zero quark chemical
potential. We know \cite{Polyakov:vu} that at high temperatures we
have center group order and at low temperatures chiral order. This
sorts for us the orientation for a possible phase structure with
respect to the condensed matter ones \cite{Calabrese:2003ia}.

Besides the temperature, which can be tuned, we also have two
independent and dynamically generated scales in the problem. The
deconfining scale $\Lambda_{\rm d}$, and the chiral symmetry
restoration scale $\Lambda_{\rm c}$. These two scales are
intimately related to the number of colors and flavors of the
theory.

However, it is the relative magnitude of these scales which is of
importance for the phase diagram. One might argue that in strong
interactions only one scale is dynamically generated. On the other
hand it is quite reasonable to imagine the dynamics driving chiral
symmetry breaking to be different than the one for center group
breaking.

\begin{figure}[h]
\includegraphics[width=4.5truecm,height=4.8truecm]{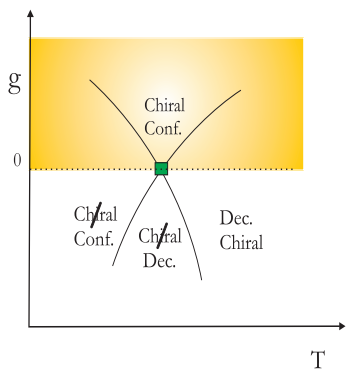}
\caption{Phase diagram displaying a tetracritical point. The
physically allowed part of the phase diagram lies beneath the
$g=0$ line.} \label{figura1}
\end{figure}

There are also theoretical arguments \cite{Casher:vw} suggesting
that $\Lambda_{\rm d}\le\Lambda_{\rm c}$ (see next section for a
more detailed discussion). It is then natural to define a new
parameter:
\begin{equation} g=\frac{\Lambda_{\rm d}-\Lambda_{\rm
c}}{\Lambda_{\rm d}} \leq 0 \ .
\end{equation}
Differently from the condensed matter cases, here $g$ cannot be
tuned but rather defines the theory. A possible phase diagram in
the $(g,T)$ plane is the one shown in fig. \ref{figura1}. We
stress that the expectation $\Lambda_{\rm d}\le\Lambda_{\rm c}$,
forces the physically allowed part of the phase diagram to lie
below the $g=0$ line.

At exactly $g=0$ tetracritical behavior would be expected, and for
this point we can translate the critical behavior discussed in the
previous sections for strong interactions. The deconfinement order
parameter symmetry fixes $N_1=1$, and we now consider different
flavors in turn.

Let us start with quenched super Yang-Mills. In this case we have
only one Majorana fermion in the adjoint representation of the
gauge group. The only global symmetry associated is an axial
symmetry which is affected by the Adler-Bell-Jackiw anomaly.
However in the quenched limit such a symmetry is restored. The
chiral symmetry is then $U(1)$ (which is also an $R$-symmetry from
the supersymmetry transformations point of view) which breaks
spontaneously. Here $N_2=2$ and if a tetracritical point would
exist it would be a biconical one. Away from the quenched limit
the $U(1)$-R symmetry is explicitly broken by an anomaly and it
might still be interesting to study what happens if one considers
this symmetry almost restorable at large $T$.

In the case of two Majorana fermions in the adjoint (i.e. one
Dirac flavor) the chiral symmetry group, after having taken into
account anomalies, is $SU(2)$, i.e. $O(3)$ with $N_2=3$. The
physics of the tetracritical point, according to high order
calculations, is the one for which the critical behaviors of the
two order parameters are unaffected by each other, i.e. we have a
decoupled fixed point. Finally for two Dirac flavors we have
$N_2=6$ and again a decoupled fixed point is expected.

Lattice simulations can determine how far we are from the
tetracritical point. {}For two colors with fermions in the
adjoint, there are numerical computations \cite{Kogut:1985xa}
which indicate that the chiral and deconfinement phase transition
happen at different temperatures. This corresponds to $g\ne 0$ and
the two transitions have the expected independent critical
behavior. It would be interesting if more recent simulations might
further investigate how competing the two orders actually are.

The main problem for not being able to reach a tetracritical point
here is that in order to change $g$ one has to change theory. In
condensed matter physics one can usually tune parameters, via
other scaling fields than the temperature, e.g. external magnetic
fields. The freedom to tune different quantities in the theory
allows, on one hand, to test the theory of critical phenomena and
to shape our understanding of phase transitions, on the other.

Since the parameter we have defined for the adjoint QCD is
dimensionless, one would expect it to be proportional to some
combination of number of colors and flavors. Then, in numerical
experiments, it might be possible to use e.g. number of flavors,
$N_f$, as a scaling field. Tuning the value of $N_f$ would affect
the relative magnitude of $\Lambda_d$ and $\Lambda_c$ and allow,
perhaps, the two transitions to close on each other. The existing
numerical investigations \cite{Kogut:1985xa} show the strong
dependence on the number of flavors for the chiral phase
transition. As already emphasized, it would be interesting to have
an up to date study of these matters.

We shall shortly see how we can achieve a multicritical point in
strong interactions with diquark condensation and confinement as
competing orders by introducing a more practical scaling field
into the problem, i.e. the quark mass.

\section{Deconfinement-Chiral Symmetry Tetracritical Point}
\label{finitemu}

In this section we investigate in some detail the two color gauge
theory with one Dirac fermion in the adjoint representation of the
gauge group. This is a theory with a number of fascinating
properties. A relevant one, for our purposes, being that when
adding a nonzero quark chemical potential one observes, at
sufficiently large baryon chemical potential, a color superfluid
transition rather than a color superconductive one
\cite{Kogut:2000ek}. This is so since we have some goldstone
bosons (pions) carrying baryonic charge.

We have divided this section into a number of subsections to help
the reader concentrate on one problem at the time, and to build
up the relevant knowledge. We will first describe the symmetries
of the fermionic action of the underlying theory, and then explore
the symmetry breaking pattern first at zero temperature and
baryon-chemical potential of the theory. We briefly review the
temperature (zero-baryon chemical potential) phase transition
scenario, which has essentially been studied in the previous
section. Subsequently, we describe the deconfining phase transition
at nonzero temperature and baryon chemical potential, while
ignoring the possible superfluid phase transition. We then
describe the superfluid phase transition at nonzero temperature
and quark chemical potential neglecting the deconfining phase
transition. We will consider both transitions simultaneously in
the next section. It is important to observe that both, the
introduction of the chemical potential as well as the presence of
a Dirac mass for the theory break explicitly the underlying global
$SU(2)$ symmetry group while preserving the $U(1)$ baryon symmetry
of the theory, as we will explicitly see below. At nonzero
temperature and nonzero baryon chemical potential we will then consider
only the exact symmetries of the problem, i.e. the center group
and the $U(1)$ symmetries.

It interesting to note, that when this theory has been
investigated in the literature at finite temperature and chemical potential,
so far attention has been paid only to the global symmetry of the theory.

\subsection{Symmetries of the Underlying Theory}
Consider one massless Dirac flavor in the adjoint representation
of two colors. The flavor group is $SU(2)$ which spontaneously
breaks to $O(2)$. The latter is the conserved quark number. In
order to elucidate all of the symmetries of the problem in detail
we write the underlying tree level Lagrangian for the fermionic
part \footnote{We use the notation of Wess and Bagger
\cite{Wess:cp} for the spinors but with metric mostly minus.} in
presence of the mass term and quark chemical potential:
\begin{eqnarray}
 + i\,\bar{Q}^A\bar{\sigma}^{\mu}D_{\mu}^{AB}Q^B -
\mu\,\bar{Q}^A\bar{\sigma}^0\,B Q^A
-\frac{m}{2}\left[Q^A\tau^1\,Q^A + h.c.\right].\nonumber
\end{eqnarray}
Here $D_{\mu}^{AB}Q^B = \partial_{\mu}\delta^{AB}Q^B - i
f^{ABC}\,G_{\mu}^B\,Q^C$, and $f^{ABC}$ are the structure
constants of the gauge group. The matrices $\tau^a$ are the pauli
matrices with the baryon number $B=\tau^3$ acting in the flavor
space, and $A=1,2,3$ is the gauge index for the fermions in the
adjoint representation. The Weyl spinor $Q^A_{\alpha,f}$, with
$\alpha=1,2$ the spin index and $f=1,2$ the flavor index, can be
represented as a vector as follows:
\begin{eqnarray}
Q^A_{\alpha}=\left(%
\begin{array}{c}
  \chi^A_{\alpha} \\
  \psi^A_{\alpha} \\
\end{array}%
\right),
\end{eqnarray}
while in the Dirac representation we have
\begin{eqnarray}
\Psi_D^A =\left(%
\begin{array}{c}
  \chi^A_{\alpha} \\
  {\bar{\psi}^{\dot{\alpha}\,A}} \\
\end{array}%
\right) .
\end{eqnarray}
At zero quark mass and chemical potential the $SU(2)$ symmetry is
evident. The extra classical $U_A(1)$ symmetry is anomalous. The
baryon number here is the $\tau^3$ generator of $SU(2)$. At non
zero baryon chemical potential and nonzero Dirac quark mass the
baryon symmetry is the only symmetry left unbroken at the
fundamental level.

\subsection{Chiral symmetry breaking: no anomaly matching but entropy-counting}
We set, for the moment, the fermion mass term and the baryon
chemical potential to zero. Usually one of the powerful methods to
discover if, in strongly interacting gauge theories, a global symmetry breaks
at low energies, is to require the global anomaly
matching conditions \cite{'tHooft:xb} among the ultraviolet and
the infrared realization of the theory. Unfortunately, for this
theory the global anomalies vanishes, since the flavor group is
$SU(2)$, and hence we cannot invoke the 't Hooft anomaly matching
conditions \cite{'tHooft:xb} to suggest that chiral symmetry must
break at low temperatures. Indeed, we can well imagine a low
temperature phase in which chiral symmetry is not broken. Although
in principle we do not need massless composite fermions, the
simplest fermions we can construct are composite objects of the
type $\lambda_{\alpha,F}\sim\bar{Q}^{\dot{\alpha}\,A}_{F}
\bar{\sigma}^{\mu}_{\dot{\alpha}\alpha}G^A_{\mu}$. Due to the
Vafa-Witten theorem \cite{Vafa:tf}, vector symmetries cannot break
spontaneously, which, in turn, means that the fermions do not
develop dynamically generated Majorana masses. However, a Dirac
mass term is of the form
$\lambda^{\alpha}_{F=1}\lambda_{\alpha\,F=2}$ and breaks the
global $SU(2)$ symmetry to the baryon number $U(1)$.

Therefore, in absence of 't Hooft anomaly matching conditions two
possible scenarios arise: We can either have spontaneous chiral
symmetry breaking, with associated two Goldstone bosons, or chiral
symmetry intact but a massless composite Dirac fermion. This is
very similar to the case of ordinary QCD with two flavors.
According to the guide suggested in \cite{Appelquist:2000qg}, the
most likely phase in the infrared is the one for which the degrees
of freedom counted according to the entropy factor, $f=\sharp~{\rm
Real~Bosons} + (7/4)\sharp~{\rm Weyl~Fermions}$, are minimized.
Here, the spontaneously broken phase has $f=2$ and the chiral
symmetry preserving phase has $f=7/2$. Chiral symmetry, here the
$SU(2)$, is therefore predicted to break at low temperatures.
Clearly these results do not depend on the number of colors. In
the case of larger number of fermion flavors the 't Hooft anomaly
conditions are non trivial and single out the infrared phase in
which chiral symmetry is broken. 't Hooft anomaly conditions have
been generalized, first, at nonzero temperature \cite{Itoyama:up}
and more recently at nonzero quark chemical potential
\cite{Sannino:2000kg}.

A possible scalar condensate must be of the form:
\begin{eqnarray}
\epsilon^{\alpha\,\beta}\langle
Q^A_{\alpha,f}\,Q^A_{\beta,f^{\prime}} \rangle\propto
E_{ff^{\prime}} \ .
\end{eqnarray}
The subgroup which leaves the condensate invariant is given by the
generators of $SU(2)$ satisfying the condition:
\begin{eqnarray} \tau^a\,E + E\,{\tau^a}^T=0 \ .\end{eqnarray}
Since the condensate is symmetric in color and antisymmetric in
spin, it must be symmetric in flavor (i.e. $E=E^T$). Requiring the
$SU(2)$ symmetry to break to its maximal orthogonal subgroup (i.e.
$O(2)$) \footnote{If the symmetry would not break to its maximal
subgroup, more goldstones would appear. These infrared realizations
would  then be disfavored, according to the entropy guide, with
respect to the case in which the $SU(2)$ breaks to the maximal
diagonal subgroup.}, we can have, for example, $E$ proportional to
the two by two identity matrix or to $\tau^{1}$. If we choose the
identity, then the unbroken generator is $\tau^2$, but if we
choose $\tau^1$, then the unbroken generator is $\tau^3$. Since we have
identified the $O(2)$ generator corresponding to the baryon number
with $\tau^3$, the condensate must be proportional to
$\tau^1$, i.e.:
\begin{eqnarray}
\epsilon^{\alpha\,\beta}\langle
Q^A_{\alpha,f}\,Q^A_{\beta,f^{\prime}}\rangle \propto
\tau^1_{ff^{\prime}}\ .
\end{eqnarray}
Two Goldstone bosons are present and are associated to the
generators $X^a=\tau^a/2$ with $a=1,2$. Note, that since the pions
here are associated to the generators which do not commute with
the baryon generator $\tau^3$ they are automatically charged under
the baryon number. The low energy effective theory in absence of
quark chemical potential is:
\begin{eqnarray}
{\cal L}_{eff}= F^2_{\pi}{\rm
Tr}\left[\partial_{\mu}U^{\dagger}\partial^{\mu}U\right] +
F^2_{\pi}m^2_{\pi}{\rm Tr}\left[U+U^{\dagger}\right] \ .
\end{eqnarray}
with
\begin{eqnarray}
U=e^{i\frac{\pi^aX^a}{F_{\pi}}}\ , \qquad a=1,2 \ ,
\end{eqnarray}
where we have introduced also a Dirac mass $m$ in the underlying
theory. Such a mass appears in the effective Lagrangian as a nonzero
mass for the pions, and one expects $m^2_{\pi}\propto m$. $U$
transforms as $g\tau^1\,Ug^T$ for $g\in SU(2)$. The previous
effective Lagrangian still preserves the $U(1)$ baryon symmetry.

Besides chiral symmetry we also have deconfinement. Here the order
parameter is the Polyakov loop, which is associated to the center
group symmetry $Z_2$ for two colors. Note that the previous
analysis is completely independent on the number of colors, which
becomes a relevant parameter only when considering the center
group symmetry as well.

\subsection{The temperature driven phase transition}
We have discussed the nonzero temperature case in section
\ref{temperature}. Here we recall the salient information needed
when endowing the quarks with a nonzero mass and chemical
potential. At zero quark chemical potential, the $SU(2)$ symmetry
is restored at a given temperature $T_{c}$, while the $Z_2$
deconfinig phase transition is indicated with $T_d$. The latter is
expected to be somewhat lower than $T_c$. If the two phase
transitions are independent, no tetracritical point is expected to
occur in this case. As soon as we add a quark mass, we expect a
cross over behavior for the $SU(2)$ phase transition. This is true
also at nonzero chemical potential, since both the mass term and
the chemical potential term explicitly break the $SU(2)$ global
symmetry. It is also worth emphasizing again, that at zero quark
chemical potential and quark mass, and due to the absence of the
't Hooft anomaly conditions to satisfy, in principle, a chiral
symmetry restoring phase transition before deconfinement might
have been possible. However, this is not allowed according to the
guide in \cite{Appelquist:2000qg}, which selects the chiral
symmetry breaking confined phase as the preferred ground state
even in absence of 't Hooft anomaly conditions. Summarizing, the
$SU(2)$ symmetry is always broken at nonzero baryon chemical
potential and Dirac mass. If a crossover phenomenon exists, it is
expected to happen, for fixed chemical
potential and quark mass, at a temperature larger or at most equal to the
critical temperature for deconfinement. As we increase the chemical
potential, the explicit breaking of the $SU(2)$ symmetry becomes
severe. We will then neglect the $SU(2)$ symmetry and analyze the
fate of the $U(1)$ baryon symmetry, the only global symmetry left
unbroken.

\subsection{The $U(1)$ baryon superfluid phase transition at nonzero $\mu$ and $T$}
As we increase the baryon chemical potential the $U(1)$ baryon
symmetry may break spontaneously. In QCD with three massless
quarks in the fundamental representation the breaking is due to a
cooper pairing phenomenon, i.e. color superconductivity.

\begin{figure}[hb]
 \includegraphics[width=7 truecm, clip=true]{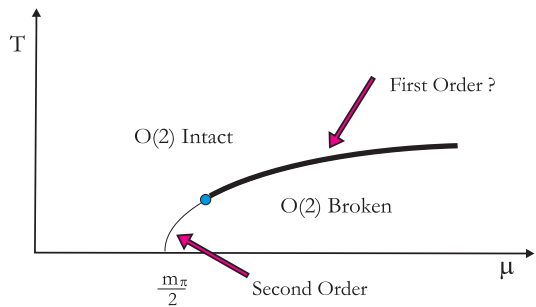}
 \caption{A schematic $(T,\mu)$-phase diagram when only the diquark condensation
 is considered}
 \label{chiral}
\end{figure}

{}For adjoint QCD the situation is different. The spontaneous
breaking of the $U(1)$ baryon symmetry is a superfluid phenomenon
\cite{Kogut:2000ek}. This is so since the pions, in this theory,
are charged under the baryon number. We have already proven this
statement in subsection B. Actually they have baryon number two
with respect to the quarks, which we have defined to have unit
baryon number. One can easily show that the chemical potential
couples directly to the pions via:
\begin{eqnarray}
\partial_{0}U\rightarrow D_0U=\partial_0 - i \mu\,
\left[U,B\right] \ .
\end{eqnarray}
After having substituted this covariant derivative in the
effective Lagrangian, a negative mass squared term proportional to
$\mu^2$ is induced. {}For $\mu> m_{\pi}/2$ the $U(1)\sim O(2)$
breaks spontaneously. On general grounds we expect two regions on
the phase diagram, one with intact $O(2)$ and the other where
$O(2)$ is spontaneously broken. This is schematically represented
in figure \ref{chiral}. The second order line starts at
$m_{\pi}/2$ at zero $T$. In literature it is argued, by computing
the effective action within the chiral perturbation theory
approach \cite{Splittorff:2002xn}, that such a second order line
ends in a tricritical point, and continues as a first order line.
There is a simple way to understand why the phase transition line
must curve to the right in the $T-\mu$ plane: By
increasing the chemical potential, we effectively increase the
negative mass squared of the goldstone boson. On the other hand,
the temperature contribution to the mass of the goldstone boson is
positive and tries to compensate the negative contribution of the
chemical potential to the squared mass term. The larger is the
chemical potential, the higher must also the temperature be to restore
the symmetry. This is, in a nutshell, the relativistic
Bose-Einstein condensation phenomenon pioneered by Haber and
Weldon in \cite{Haber:1981ts}.

Both the critical temperature and the critical chemical potential
of the tricritical point increase with the pion mass
\cite{Splittorff:2002xn}. What is relevant for us is that: i) two
well separated regions exist, and ii) we have a second order phase
transition near $\mu=m_{\pi}/2$.

\subsection{Deconfinement at nonzero $\mu$ and $T$.}
\begin{figure}[h]
\includegraphics[width=6.3 truecm, clip=true]{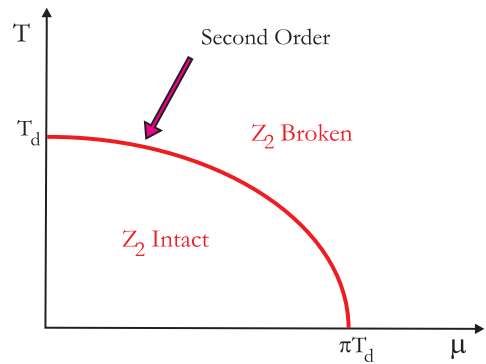}
\caption{A schematic $(T,\mu)$-phase diagram where only
deconfinement is considered} \label{Deconfinement}
\end{figure}

As already stated, the presence of quarks in the adjoint
representation of the gauge group does not break the center group
symmetry. Note also, that up to now the color played little role.
In other words, whether the center group is $Z_2$ or $Z_3$, one
expects the chiral symmetry part of the analysis (here also the
$U(1)$ baryon symmetry is termed chiral symmetry) to be to a
large extent unaffected. This, however, is not true, as we will
demonstrate below. In this subsection we only consider the pure
deconfinement phase transition.  Two distinct regions in the phase
diagram occur: in one we have center group order (i.e.
deconfinement) and in the other we have disorder (i.e.
confinement). If the number of colors is larger than two we expect
a first order line, while if the number of colors is two, a second
order line is most likely to occur. Let us consider the two color
case: Then a possible phase diagram (for deconfinement only) is
provided in figure \ref{Deconfinement}.
We have not considered the possibility of a tricritical point, but
here the important point is that there are two well separated
regions. We have simply estimated the critical chemical potential
for deconfinement to be of the order of $\sim \pi T_{d}$, with
$T_d$ the deconfinement temperature at zero chemical potential.
This value is meant only to guide our intuition, and it has been
obtained using the bag model. However, we do expect the correct
value to be near the one predicted. More specifically, the
contributions to the pressure from free gluons and quarks in the
adjoint representation are, respectively,
\begin{eqnarray}
P_g &=& g_g\frac{\pi^2 T^4}{90}, \\
P_q &=& g_q T^4 \left [
\frac{7\pi^2}{180}+\frac{1}{6}\frac{\mu^2}{T^2}+\frac{1}{12\pi^2}\frac{\mu^4}{T^4}
\right ],
\end{eqnarray}
where generally $g_g=(N_c^2-1)$ and $g_q=N_f (N_c^2-1)$, and we
set $N_c=2$ and $N_f=1$. The phase transition line in the
$(T,\mu)$-plane is determined through \begin{equation} P_g+P_q=B,
\end{equation} where $B$ is the bag constant. We determine $B$ at
zero chemical potential, and using the value so obtained, we find
at zero $T$ that $\mu_{\rm d}=0.9\pi T_{\rm d}$ for the
deconfinement transition. Ultimately this value will have to be
determined via lattice simulations. The above computation is meant
to be just a rough estimate.

If we take the number of colors larger than two, the second order
deconfinement line is replaced by a first order one.

\section{Emergence of a tetracritical point}
\label{tetracritical} 
\begin{figure}[h]
\includegraphics[width=7.5 truecm, clip=true]{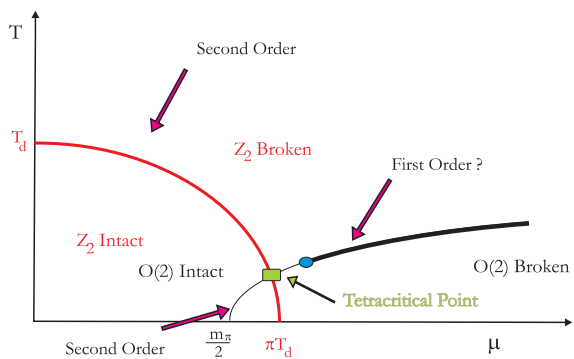}
\caption{A possible $(T,\mu)$-phase diagram when both
 possible phase transitions, chiral and deconfinement, are taken into account.}
\label{Tetra}
\end{figure}

The previous analysis neglects the fact that
the two order parameters (i.e. the Polyakov loop and the diquark
condensate) can and will compete. To argue that a tetracritical
point is a natural outcome, take the pion mass to be lighter than
twice the critical chemical potential (near zero temperature) for
deconfinement, $m_{\pi}\lesssim 2\pi\,T_{d}$. Now the two curves,
i.e. the one for deconfinement and the one for the $U(1)$ baryon
(or chiral) symmetry breaking, meet at a tetracritical point as
qualitatively illustrated in the figure \ref{Tetra}.
We have chosen, to plot the curves, the pion mass to be such that
the tetracritical point occurs when the two second order lines
meet. A tetracritical point is a very intriguing possibility and
the two order parameters here will influence each other. So, the
naive expectation that in the adjoint representation chiral
symmetry and deconfinement do not communicate is misleading.

By tuning the value of $m_\pi$ one can tune the position of the
diquark condensation line with respect to the deconfinement one.
Here the pion mass plays the role of the anisotropy parameter.

Near the tetracritical point one can apply the results of a
standard $\epsilon$ expansion analysis as discussed earlier. The
tetracritical fixed point in adjoint QCD with single Dirac flavor,
when the two second order lines meet, is in the universality class
of the $O(1)\oplus O(2)$ theory. The effective potential contains
the Polyakov loop $\ell$ and the matrix $U$ which corresponds in
practice to a complex scalar field, or two-component real field.
Due to such a group structure, using the results of
\cite{Fisher2,Calabrese:2003ia}, we predict the tetracritical
point to be a nontrivial (i.e. non decoupled) biconical one.  The
critical exponents are provided in section \ref{multicritical}.

\begin{figure}[ht]
\includegraphics[width=7.5 truecm, clip=true]{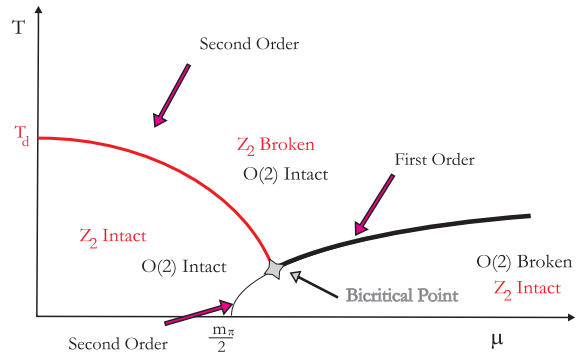}
\caption{A possible $(T,\mu)$-phase diagram when both
 possible phase transitions, chiral and confinement, are taken into account and meet
at a bicritical point.}
\label{Bicritical}
\end{figure}

Other interesting phase diagrams can be considered: {}For example,
by tuning the quark mass the first order chiral line can meet the
second order deconfinement transition. As another alternative,
while we have assumed here the deconfinement transition to be
second order over the whole $T-\mu$ plane, we cannot generally
exclude the possibility that the deconfinement line develops a
tricritical point before meeting the chiral line. Also, when the
number of colors is larger than $2$, the deconfinement line is
always first order. We do not exclude the possibility that for
similar theories one could observe the appearance of a bicritical
point. In this case a typical phase diagram is depicted in figure
\ref{Bicritical}.
If the pion mass is sufficiently large, deconfinement is expected
to occur before spontaneous breaking of the baryon number. In this
regime the two order parameters do not compete anymore.
Clearly all of these possibilities are
intriguing and deserve to be investigated.

\section{Conclusions and suggestions}
\label{conclusions}
We have shown that when the fermions are in
the adjoint representation of the gauge group, a tetracritical
fixed point naturally emerges. This is possible since the $Z_N$
symmetry associated with deconfinement is well defined in this
theory. The tetracritical point lies in the $T-\mu$ plane and for
two colors may be biconical with a suitable choice of the quark
mass. What is interesting, is that in this way we can
quantitatively test the effects of confinement, or center group
symmetry, on a chiral symmetry type phase transition and vice
versa.

{}For quarks in the fundamental representation of the gauge group
the possibility of a tetracritical point is not expected, since
the center group symmetry is explicitly broken. Besides, the
breaking of the $Z_N$ symmetry was used to explain in
\cite{Mocsy:2003qw,{Mocsy:2003tr}}, via a simple effective
Lagrangian, how deconfinement and chiral symmetry are intertwined
not only at the level of susceptibilities but also at the level of
condensates. The results in our earlier investigations were able
to provide a general qualitative understanding of the lattice
data. It is, however, still possible, although unlikely (see the
discussion in \cite{Mocsy:2004yt}), that the breaking of the
center group symmetry (due to the quarks in the fundamental
representation of the center group symmetry) is dynamically
suppressed. Such a breaking is much attenuated, for example, when
considering a small ratio of the number of flavors over the number
of colors. If such a dynamical suppression of the center symmetry
breaking occurs in the chiral limit, a (quasi)tetracritical point
may be observed in lattice simulations. Unfortunately, it is very
hard to disentangle such a behavior if the phase transitions are
of first order, and hence this behavior might be better tested in
two color QCD with one Dirac flavor or two Dirac flavors in the
fundamental representation. The tetracritical point on the
temperature axis would be characterized by a $O(1)\oplus O(3)$ or
$O(1)\oplus O(6)$ symmetry respectively. A decoupled tetracritical
point would emerge with independent Ising and Heisenberg
behaviors. Considering this scenario at any nonzero quark masses,
the $O(1)$ symmetry would be (quasi)exact, and the chiral
transition would be then induced \cite{Mocsy:2003qw}. The critical
exponents are well known here. Departures from these limiting
behaviors is a measure of the amount of center symmetry breaking
induced by the presence of the quarks in the fundamental
representation of the gauge group.

\acknowledgments We thank P.H. Damgaard, K. Rummukainen and K.
Splittorff for careful reading of the manuscript. We acknowledge
useful discussions with A.D. Jackson, K. Kajantie, A. M\'{o}csy,
R. Pisarski and B. Svetitsky.


\end{document}